\documentclass[preprintnumbers,showpacs,amsmath,amssymb,floatfix,prd,twocolumn,superscriptaddress,nofootinbib]{revtex4}
\usepackage{graphicx}
\usepackage{epsfig}
\usepackage{bm}
\usepackage{amsfonts}
\usepackage{epstopdf}
\usepackage{bm}

\begin{document}

\title{ Towards a realistic solution of the cosmological constant fine-tuning problem \\by Higgs inflation}

\author{Chao-Jun Feng}
\email{fengcj@shnu.edu.cn} 
\affiliation{Shanghai United Center for Astrophysics (SUCA), \\ Shanghai Normal University,
    100 Guilin Road, Shanghai 200234, P.R.China}
    \affiliation{State Key Laboratory of Theoretical Physics, \\Institute of Theoretical Physics, Chinese Academy of Sciences, Beijing 100190, P.R.China}

\author{Xin-Zhou Li}
\email{kychz@shnu.edu.cn} \affiliation{Shanghai United Center for Astrophysics (SUCA),  \\ Shanghai Normal University,
    100 Guilin Road, Shanghai 200234, P.R.China}

\begin{abstract}
Why  the cosmological constant $\Lambda$ observed today is so much smaller than the Planck scale or why the universe is accelerating at present? This is  so-called the cosmological constant fine-tuning problem. In this paper, we find that this problem is solved  with the help of Higgs inflation by simply assuming  a variable cosmological ``constant" during  the inflation epoch. In the meanwhile, it could predict a large tensor-to-scalar ratio $r\approx 0.20$ and a large running of spectral index $n'_s \approx -0.028$ with a red-tilt spectrum $n_s \approx 0.96$, as well as a big enough number of e-folds $N\approx 40$ that required to solve the problems in the Big Bang cosmology with the help of $\Lambda$.
\end{abstract}

\maketitle


\section{Introduction}

The detection of B-mode from CMB by the BICEP2 group \cite{Ade:2014xna} has indicated a strong evidence of inflation \cite{Guth:1980zm,Linde:1981mu,Albrecht:1982wi}.  According to the results of BICEP2, the tensor-to-scalar ratio is constrained to $r = 0.20_{-0.05}^{+0.07}$ at  $68\%$ CL for the lensed-$\Lambda$CDM model, with $r=0$ disfavoured at $7.0\sigma$ level.  With the help of this new constant on $r$ as well as that on the spectral index, some inflation models with prediction of negligible tensor perturbation have been excluded, such as the small-field inflation models. 

In a simplest slow-roll inflation model, the early universe was driven by a scalar field called the inflaton. To achieve an successful inflation, the potential of the inflaton field should be very flat. The most economical and fundamental candidate for the inflaton is the standard model (SM) Higgs boson, which has been already observed by the collider experiment LHC in 2012 \cite{Aad:2012tfa,Chatrchyan:2012ufa}.  In this sense, Higgs inflation is a simple and elegant model. However, to realize an inflation model by the Higgs boson is not easy to predict  correct density perturbations. To see this, one could simply estimate the mass of the inflaton from the the amplitude  of the scalar perturbation power spectrum: $m\approx 1.5\times 10^{13}$ GeV. which is many orders of magnitude larger than the observed Higgs mass, $m_{h}\approx 125.9\pm0.4$ GeV. In other words, the potential of Higgs field $h$ is not flat enough to realize an inflation.

By introducing a non-minimal coupling to the gravity ($\sim h^2 R$) ,  one could indeed achieve such a flat potential  after a conformal transformation \cite{Bezrukov:2007ep}. But this model can not accommodate the new measurement from BICEP2, since it generally predicts a small amplitude of tensor perturbations, see ref.\cite{Cook:2014dga}. Some other alternative Higgs inflation models are summarised in ref.\cite{Feng:2014naa} and the references therein. Among these models, there is a very interesting Higgs inflation model called the Higgs chaotic inflation  proposed in ref.\cite{Nakayama:2010sk}. In this model, the quadratic chaotic inflation model could be realized by the SM Higgs boson  based on the so-called running kinetic inflation \cite{Nakayama:2010kt, Takahashi:2010ky,Nakayama:2014koa}.  We will give a very brief review on this model in the next section.

Recently, it is found that the cosmological constant plays an important role during the inflation time \cite{Feng:2014naa}. Not only the tensor-to-scalar ratio could be enhanced   with the help of the cosmological constant, but also the number of e-folds could be  large enough to overcome the problems in the Big Bang cosmology. 

However, there are still two problems left. One is the famous cosmological constant fine-tuning problem, which states why the cosmological constant observed today is so much smaller than the Planck scale, since the effective field theory still valid near this high energy scale.  By using the observational value of the Hubble constant~\cite{Ade:2013uln}: $H_0\approx 2.1332h \times 10^{-42} $GeV  with $h\approx 67$, and the value of the energy density parameter of the cosmological constant \cite{Ade:2013uln}: $\Omega_{\Lambda0}\approx0.68$, one can estimate that $\Lambda_0/M_{pl}^2=3H_0^2\Omega_{\Lambda0}/M_{pl}^2 \approx \mathcal{O}(10^{-117} ) \ll 1$. This is equivalent to ask why the universe is accelerating at present. To explain this later time acceleration, people usually introduce an exotic energy component with negative pressure called the dark energy \cite{Li:2012dt}, but so far as we known, we have not find it in the laboratory yet. 

The other problem is that there is a tension between the observational results from BICEP2 and the report of \textit{Planck}. According to the recent  works from \textit{Planck} \cite{Ade:2013uln}, it gives $n_s=0.9600\pm0.0071$ and $r_{0.002}<0.11$ at $95\%$CL by combination of \textit{Planck}+WP+highL data.  However, when the running parameter defined as $n'_s \equiv dn_s/d\ln k$ is included in the data fitting, it gives $n_s=0.95700\pm0.0075$, $r_{0.002}<0.26$ and  $n'_s = -0.022_{-0.021}^{+0.020}$ at $95\%$CL by the same data.  In a word, to give a consistent constraint on $r$ for the combination of  \textit{Planck}+WP+highL  data and the BICEP2 data, we require a large running of the spectral index .  

In history, the cosmological constant was first introduced by Einstein in his  field equation as follows
\begin{equation}\label{equ:ein}
	R_{\mu\nu} - \frac{1}{2}R g_{\mu\nu} + \Lambda g_{\mu\nu} = 8\pi G T_{\mu\nu}\,,
\end{equation}
where $R_{\mu\nu}$ and $R$ are the Ricci  tensor and the scalar curvature respectively, while $T_{\mu\nu}$ is the energy-momentum tensor of matters.   Using the Bianchi identities and assuming the energy-momentum tensor satisfies  the conservation law $\nabla^\mu T_{\mu\nu} =0$,  it follows that the covariant divergence of $\Lambda g_{\mu\nu}$ must vanish also, and hence that $\Lambda = $ const. However, if we move the cosmological term $\Lambda g_{\mu\nu}$ to the right-hand side of Eq.(\ref{equ:ein}), and to interpret $\Lambda$ as part of the matter content of the universe with total energy-momentum tensor $\tilde T_{\mu\nu} = T_{\mu\nu} -  \Lambda g_{\mu\nu}$. Hereafter we set the unit $8\pi G = 1$ for convenience. Once this is done, there are no a prior reasons why $\Lambda$ should not vary, as long as the total energy-momentum tensor satisfies the conservation law: $\nabla^\mu \tilde T_{\mu\nu} =0$. For a wonderful review on the variable cosmological constant, see ref.\cite{Overduin:1998zv}.

In this paper, we will also assume that $\Lambda$ is no longer a constant during the inflation time, which means there exits an energy transfer between $\Lambda(t)$ and the inflaton field, see Sec.\ref{sec:main}. And we find that the cosmological constant fine-tuning problem is solved with the help of inflation, while a large running of the spectral index is obtained with the help of the cosmological constant. In the following, we still call $\Lambda(t)$  the cosmological constant for the sake of historical convention, but one should keep in mind that $\Lambda(t)$ is time-dependent now.

\section{Running kinetic inflation}
The running kinetic inflation can be easily implemented in supergravity by assuming a shift symmetry exhibiting itself in the K\"ahler potential at high energy scales, while this symmetry is explicitly broken and therefore becomes much less prominent at low energy scales. In the unitary gauge, one can write down the Lagrangian for the Higgs boson $h$ \cite{Nakayama:2010sk, Nakayama:2010kt, Takahashi:2010ky, Nakayama:2014koa}:
\begin{equation}\label{equ:lag}
	\mathcal{L} = \frac{1}{2} \left(1+ \xi \frac{h^2}{2} \right) (\partial h)^2 - \frac{\lambda_h}{4} (h^2 - v^2)^2 \,.
\end{equation}
The effect of non-canonical kinetic term is significant for large $h\geq 1/\sqrt{\xi}$.  The kinetic term grows as $h^2$, that is why the name ``running kinetic inflation''.   By redefining the Higgs field, one can rewrite the Lagrangian in terms of canonically normalized field $\phi \equiv \sqrt{\xi/8} h^2$ with the effective potential 
\begin{equation}\label{equ:potential}
	V(\phi) = \frac{1}{2} m^2 \phi^2 \,, \quad m^2 \equiv \frac{4\lambda_h}{\xi} M_{pl}^2 \,.
\end{equation}
Thus, the quadratic chaotic inflation occurs.

\section{Variable cosmological ``constant" during inflation}\label{sec:main}

During inflation, the spatially flat universe  could be described by the  Friedmann equation $3H^2= \rho_\phi + \rho_\Lambda $, where $\rho_\phi =  (\dot\phi^2 + m^2\phi^2)/2$ is the energy density of the inflaton, which is dominated by its potential energy, while $\rho_\Lambda = \Lambda$ in the unit of $8\pi G =1$.  Furthermore, by using the slow-roll approximation, the Friedmann equation could be approximated as 
\begin{equation}\label{equ:fried}
	3H^2 = \frac{1}{2}m^2\phi^2 + \Lambda \,.
\end{equation}
Also, the conservation law $\nabla^\mu \tilde T_{\mu\nu} =0$ gives 
\begin{equation}\label{equ:con}
	\dot \rho_\Lambda + \dot \rho_\phi + 3H(\rho_\phi + p_\phi) = 0\,,
\end{equation}
with the pressure density of inflaton field, $p_\phi =  (\dot\phi^2 - m^2\phi^2)/2$. Notice that the time evolution of $\rho_\Lambda$ and $\rho_\phi$ is coupled in Eq.(\ref{equ:con}), and there is an energy transfer between the inflaton and $\Lambda$. In fact, we have $p_\phi \approx - \rho_\phi$ during inflation time. Then according to  conservation law (\ref{equ:con}), it follows that $\dot\rho_\phi = - \dot\rho_\Lambda$, which means the energy could be transformed from $\Lambda$ to $\rho_\phi$ if $\dot\rho_\Lambda <0$.  In other words, there exists an interaction between $\Lambda$ and the inflaton field, which causes the transfer of energy but keeps the total energy density unchanged, since $\dot \rho_\Lambda + \dot\rho_\phi = 0$.

The simplest term in the Lagrangian that describing the interaction  between the two components might be  $\mathcal{L}_{\rm int} \sim \phi \rho_\Lambda$, which is analogous to the particle-fluid interactions that often emerged in the particle physics. Hence, the conservation law becomes
\begin{eqnarray}
	\dot \rho_\phi+ 3H(\rho_\phi + p_\phi)  &=& Q \,, \label{equ:eom1}\\
	\dot \rho_ \Lambda &=& -Q \,, \label{equ:eom2}
\end{eqnarray}
with  the coupling term 
\begin{equation}\label{equ:q1}
	Q = -\alpha  \dot\phi  \rho_\Lambda= -\alpha  \dot\phi  \Lambda \,,
\end{equation}
where $\alpha >0$ is a dimensionless constant. Here one can see that, the interaction term $Q \neq 0$ as long as the inflaton has non-vanishsed kinetic energy $\dot\phi\neq0$, and $Q$ would be almost vanished when $\Lambda$ becomes extremely small.  By inserting the explicit form of $\rho_\phi$ into Eq.(\ref{equ:eom1}), we obtain 
\begin{equation}\label{equ:eom11}
	\ddot\phi + 3H\dot\phi + m^2\phi + \alpha \Lambda = 0\,,
\end{equation}
and in the slow-roll limit, it becomes
\begin{equation}\label{equ:eom12}
	\dot\phi \approx  - \frac{m^2\phi +\alpha \Lambda}{ 3H} <0 \,.
\end{equation}
Thus,  we have a positive interaction term  $Q>0$, which means that the energy density of the cosmological constant $\rho_\Lambda$ is decreasing while  that of the inflaton $\rho_\phi$ is increasing. 

From Eq.(\ref{equ:eom2}), we have $\dot\Lambda= \alpha  \dot\phi \Lambda$. Therefore, we obtain
\begin{equation}\label{equ:cc}
	\Lambda_f
	=     \Lambda \exp{\bigg[- \alpha(\phi - \phi_f)  \bigg]} \approx \Lambda\exp{\bigg(- \alpha \phi \bigg)} \,,
\end{equation}
where the subscript $f$ denoted the values at the end of inflaton. It should be noticed that Eq.(\ref{equ:cc}) is very important to explain why the cosmological constant is so small at present universe with the help of inflaton field $\phi$. In the following, one can see that $\Lambda_f$ could be indeed as the same order of magnitude as its present value, namely, $\Lambda_f \approx \Lambda_0$, which is extremely small. Then the interaction term in Eq.(\ref{equ:q1}) is almost vanished, $Q\approx0$ or $\dot \Lambda \approx 0$.  Therefore, $\Lambda_f$ becomes a constant after inflation, and  it will cause a second accelerating at later time that has been confirmed by  the supernovae observations.  

On the other hand, with the help of $\Lambda$  and the coupling between  the inflation and itself, the inflation model could predict a large tensor-to-scalar ratio $r\approx 0.2$ and a large running of the spectral index $n_s'\approx -0.03$ with $n_s \approx 0.96$ and the number of e-folds $N \approx 40$. To see this,  one may define the following slow-roll parameters
\begin{eqnarray}
	\epsilon &\equiv& -\frac{\dot H}{H^2} =  \frac{ 2(m^2\phi +\alpha \Lambda)^2}{ (m^2\phi^2 + 2\Lambda)^2} \,, \label{equ:slowroll1}\\
	\eta &\equiv& \frac{\dot \epsilon}{\epsilon H}  = -4 \frac{ m^2   +\alpha^2 \Lambda }{ m^2\phi^2 +2 \Lambda} +  4\epsilon \,.\label{equ:slowroll2}
\end{eqnarray}
It is clearly that when the field value $\phi$ is large, the slow-roll parameters are small enough, i.e. $\epsilon, \eta \ll 1$,  to drive an inflation. And the inflation will end when  $\Lambda$ goes small and  $\phi_f \lesssim 1$ , namely, $\epsilon \approx 2/\phi^2_f \approx 1$.

Furthermore, the power spectrum of the scalar perturbation is given by
\begin{equation}\label{equ:ps}
	\mathcal{P}_s = A_s \left( \frac{k}{k_*}\right)^{n_s-1+ n_s'\ln(k/k_*)/2} \,,
\end{equation}
with  the amplitude $A_s$  given by 
\begin{equation}\label{equ:amp}
		A_s = \frac{H^2}{8\pi^2 \epsilon}  \approx \frac{m^2\phi^2+2\Lambda}{3\pi^2  r} \,,
\end{equation}
where we have used the Friedmann equation (\ref{equ:fried}) and also the relation $r=16\epsilon$.  Here, $n_s$ is called the spectral index, and $n_s' =  dn_s/d\ln k$ is the running of the index. And also the spectral indices and its running could be given in terms of the slow-roll parameters:
\begin{eqnarray}
	n_s - 1 &\approx&  -2\epsilon - \eta  \,,  \label{equ:ns}\\
	n_s'  &\approx& - 2\alpha^3 \Omega_\Lambda \sqrt{2\epsilon} + 8\epsilon^2-  8\epsilon\eta\,, \label{equ:run}
\end{eqnarray}
where we have defined $\Omega_\Lambda = \Lambda/(3H^2)$. Notice that, the interaction is also important to produce a relatively large running of the spectral index trough the first term in Eq.(\ref{equ:run}). Without interaction, i.e. $\alpha=0$, one could only get a small running of order $\mathcal{O}(\epsilon^2)$.  In fact, one can rewrite the running of spectral index as the following
\begin{equation}\label{equ:run2}
	n'_s  = - 2\alpha^3 \Omega_\Lambda \sqrt{\frac{r}{8}} +\frac{r}{2}\left(n_s - 1 +  \frac{3r}{16}\right) \,.
\end{equation}

\begin{figure}[h]
\begin{center}
\includegraphics[width=0.45\textwidth,angle=0]{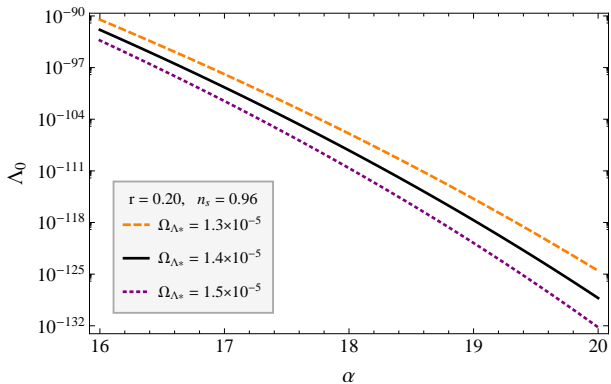}
\caption{\label{fig:la}The present value of  $\Lambda_0$ (measured in the unit of $M_{pl}$.) v.s.  the parameter $alpha$.  The dashed-orange, solid-black, and  dotted-purple curves correspond to $10^5\Omega_{\Lambda*} = 1.3, 1.4, 1.5$ respectively.  Here the tensor-to-scalar ratio and the scalar power spectral index are fixed at $r=0.20$ and $n_s=0.96$, respectively.    }
\end{center}
\end{figure}

The inflaton mass and its field value could be also obtained in terms of $n_s, r, A_s$ and $\Omega_\Lambda$ as 
\begin{eqnarray}
	m^2 &=& \frac{{3\pi^2 r A_s} }{4} \left(n_s - 1 +\frac{3r}{8} -  2\alpha^2 \Omega_\Lambda \right) \,, \label{equ:mass} \\
\nonumber
	\phi &=& 2\left( \sqrt{\frac{r}{8}} - \alpha \Omega_\Lambda \right) \left(n_s - 1 +\frac{3r}{8} -  2\alpha^2 \Omega_\Lambda \right)^{-1} \,,\label{equ:phi}\\
\end{eqnarray}
where $\Omega_\Lambda$ satisfies the Friedmann equation  (\ref{equ:fried}) that be rewritten as  
\begin{equation}\label{equ:maineq}
	\left( \sqrt{\frac{r}{8}} - \alpha \Omega_\Lambda \right)^2 \left(n_s - 1 +\frac{3r}{8} -  2\alpha^2 \Omega_\Lambda \right)^{-1}  = (1-\Omega_\Lambda)\,.
\end{equation}
By using Eq.(\ref{equ:cc}) and the value of inflaton field (\ref{equ:phi}), we obtain the following equation
\begin{widetext}
\begin{equation}\label{equ:maineq2}
	 \left(n_s - 1 +\frac{3r}{8} -  2\alpha^2 \Omega_\Lambda \right)
	 \bigg[ \ln \left(  \frac{2\Lambda_0}{3\pi^2  A_s}\right) - \ln \left( r  \Omega_\Lambda \right) \bigg]
	 = -2\alpha \left( \sqrt{\frac{r}{8}} - \alpha \Omega_\Lambda \right) \,.
\end{equation}
\end{widetext}
The number of e-folds before the ending of inflation is defined as
\begin{equation}\label{equ:efold}
	N\equiv  \int_{t_*}^{t_f} H dt = \int_{\phi_*}^{\phi_f} \frac{H}{\dot\phi} d\phi =  - \frac{1}{2}\int _{\phi_*}^{\phi_f} \frac{ m^2\phi^2 + 2\Lambda }{  m^2\phi +\alpha \Lambda} d\phi \,,
\end{equation}
where $\phi_*$ is the value of the inflaton field at time $t_*$  when there are $N$ e-foldings to the end of inflation.  Substituting the observational values of $r\approx0.2, n_s\approx 0.96, A_s \approx 2.19\times 10^{-9}$ and $\Lambda_0 = 3H_0^2\Omega_{\Lambda0} \approx  7.054 \times 10^{-117} $ into Eqs.(\ref{equ:maineq}) and (\ref{equ:maineq2}), we estimate the values of $\alpha \approx 18.74$ and $\Omega_{\Lambda *} = \Omega_\Lambda(t_*) \approx 1.436\times10^{-5}$.  Then the number of e-folds before the ending of inflation is given by $N\approx 40$, while the corresponding value of inflaton is $\phi_* \approx 12.67 M_{pl}$. It indicates that the inflation field dominates the universe when there are about $40$ e-foldings to the end of inflation, while it is also accompanied by a small but very important cosmological constant component. The energy scale of cosmological constant  at time $t_*$ is roughly $\Lambda_* \approx 9.31\times 10^{-14} M_{pl}^2 \approx (7.43\times 10^{11} \text{GeV} )^2$.   In other words, when  $\alpha$ and $\Omega_{\Lambda*}$ are around the values estimated above,  one can finally obtain a very small $\Lambda_0$ just as the same order of magnitude as its present value, see Fig.\ref{fig:la}. Thus, the cosmological constant fine tuning problem is solved by inflation!

In fact, inflation could last longer than $40$ e-foldings. Assuming the inflation begins at the time $t_i$  when value of the inflaton is $\phi_i \approx 14.27 M_{pl}$, then the corresponding energy scale of cosmological constant is just at the Planck scale $\Lambda_i \approx M_{pl}^2 \approx (2.435\times10^{18} \text{GeV})^2$. Notice that before the time $t_i$, the effective quantum field theory will break down, and some modifications from the quantum gravity theory are required.

Also, one can estimates the mass of inflaton field by substituting $\alpha$ and $\Omega_{\Lambda*}$ into Eq.(\ref{equ:mass}): $m\approx 2.19\times 10^{13}$GeV, which is slighly smaller than that in ref.\cite{Feng:2014naa}.   If $\xi$ is sufficiently large, say $\xi \approx 6.4\times 10^9$ in Eq.(\ref{equ:potential}), the quartic coupling could be  $\lambda_h \approx 0.13$, which is required to explain the correct electroweak scale and the Higgs boson mass $m_h=\sqrt{2\lambda_h} v$. The large value of $\xi$ could be understood in terms of symmetry, see refs.\cite{Nakayama:2010sk, Nakayama:2010kt, Takahashi:2010ky, Nakayama:2014koa} for details.

\begin{figure}[h]
\begin{center}
\includegraphics[width=0.45\textwidth,angle=0]{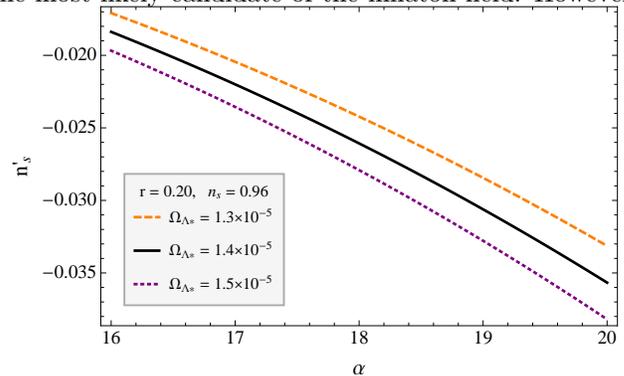}
\caption{\label{fig:na}The running of the scalar power spectral index $n'_s$ v.s.  the parameter $alpha$.  The dashed-orange, solid-black, and  dotted-purple curves correspond to $10^5\Omega_{\Lambda*} = 1.3, 1.4, 1.5$ respectively.  Here the tensor-to-scalar ratio and the scalar power spectral index are fixed at $r=0.20$ and $n_s=0.96$, respectively.    }
\end{center}
\end{figure}

The running of the spectral index can be also estimated as $n'_s \approx -0.028$, when $\alpha \approx 18.74$ and $\Omega_{\Lambda *}  \approx 1.436\times10^{-5}$. In the meanwhile, it predicts the tensor-to-scalar ratio $r\approx0.20$ and the spectral index $n_s\approx0.96$. We  also plot the values of $n'_s$ with different $\alpha$ and $\Omega_{\Lambda *}$ in Fig.\ref{fig:na}. Generally, one can see that 
the value of $n'_s$ could be as larger as the central value $n_s\approx -0.02$ from the fitting results of \textit{Planck} \cite{Ade:2013uln}.  In this sense, we have solved the large running of the spectral index problem of inflation. 

\section{Conclusion and discussion}

The recent detection of B-mode by BICEP2 indicates an exciting leap forward in our ability to explore the early universe and fundamental physics. The measurement of the tensor-to-scalar ratio $r\approx0.2$ shows a very powerful constraint to theoretical inflation models. Higgs boson is the most likely candidate of the inflaton field. However, its mass $m_h\sim\mathcal{O}(10^2)$ GeV is much smaller  than that for a inflaton $m\sim\mathcal{O}(10^{13})$ GeV. To solve this  hierarchy  problem, some authors proposed so-called the Higgs chaotic inflation model based on the running kinetic inflation. In ref.\cite{Feng:2014naa} the authors find the cosmological constant $\Lambda$ plays an important role in the Higgs inflation, but there still lefts the fine-tuning problem of $\Lambda$ and the large running of spectral index challenge  of the inflation. In this paper, we suggest a simplest coupling between $\Lambda$ and the inflaton field, $\mathcal{L}_{\rm int} \sim \phi \Lambda$ with a variable $\Lambda(t)$ to solve these two problems and we finally make it.

\acknowledgments

This work is supported by National Science Foundation of China grant Nos.~11105091 and~11047138, ``Chen Guang" project supported by Shanghai Municipal Education Commission and Shanghai Education Development Foundation Grant No. 12CG51, National Education Foundation of China grant  No.~2009312711004, Shanghai Natural Science Foundation, China grant No.~10ZR1422000, Key Project of Chinese Ministry of Education grant, No.~211059,  and  Shanghai Special Education Foundation, No.~ssd10004, and the Program of Shanghai Normal University (DXL124).

\end{document}